\documentclass[10pt,journal,twocolumn,]{IEEEtran}
\IEEEoverridecommandlockouts
\usepackage{epsfig,cite,bm,algorithm,algorithmic,epstopdf,amsmath,subfigure,multirow,amssymb,amsfonts,amsthm,xcolor,caption,graphicx,textcomp,xcolor, subfigure, multirow, array, makecell}
\usepackage[inline, shortlabels]{enumitem}
\setlist{itemjoin ={,\enspace},itemjoin* = { and\enspace}}

\usepackage[T1]{fontenc}
\usepackage{stfloats}
\allowdisplaybreaks[4]

\allowdisplaybreaks[4]

\begin{document}
	
\title{Revolutionizing Medical Data Transmission with IoMT: A Comprehensive Survey of  Wireless Communication Solutions and Future Directions}
	
	\author{Jiasi Zhou, Yanjing Sun, and Chintha Tellambura,~\IEEEmembership{Fellow,~IEEE}
		\thanks{Jiasi Zhou is with the School of Medical Information and Engineering, Xuzhou Medical University, Xuzhou, 221004, China, (email: jiasi\_zhou@xzhmu.edu.cn). (\emph{Corresponding author: Jiasi Zhou}).}
        \thanks{Yanjing Sun is with the School of Information and Control Engineering, China
        University of Mining and Technology, Xuzhou 221116, China (email: yjsun@cumt.edu.cn).}
		\thanks{ Chintha Tellambura is with the Department of Electrical and Computer Engineering, University of Alberta, Edmonton, AB, T6G 2R3, Canada (email: ct4@ualberta.ca).} 
		\thanks{This work was supported by the Talented Scientific Research Foundation of Xuzhou Medical University (D2022027).}}
	\maketitle
	
	\begin{abstract} 
		Traditional hospital-based medical examination methods face unprecedented challenges due to the aging global population. The Internet of Medical Things (IoMT), an advanced extension of the Internet of Things (IoT) tailored for the medical field, offers a transformative solution for delivering medical care. IoMT consists of interconnected medical devices that collect and transmit patients' vital signs online. This data can be analyzed to identify potential health issues, support medical decision-making, enhance patient outcomes, and streamline healthcare operations. Additionally, IoMT helps individuals make informed decisions about their health and fitness. There is a natural synergy with emerging communication technologies to ensure the secure and timely transmission of medical data. This paper presents the first comprehensive tutorial on cutting-edge IoMT research focusing on wireless communication-based solutions. It introduces a systematic three-tier framework to analyze IoMT networks and identify application scenarios. The paper examines the medical data transmission process, including intra-wireless Body Area Networks (WBAN), inter-WBAN, and beyond-WBAN communications. It also discusses the challenges of implementing IoMT applications, such as the longevity of biosensors, co-channel interference management, information security, and data processing delays. Proposed solutions to these challenges are explored from a wireless communication perspective, and future research directions are outlined. The survey concludes with a summary of key findings and insights.
	\end{abstract} 
	
	
	\section{Introduction}

	\subsection{Background and Motivations}
	According to the United Nations Population Division, the world's population aged $60$ years and older is expected to reach $1.4$ billion in $2030$ and $2.1$ billion in $2050$\cite{9241882}. The aging population imposes unprecedented pressure on the regular hospital-based approach to medical examinations. The phenomenon became more apparent during the COVID-19 pandemic, which pushed many medical centers to breaking point\cite{SWAYAMSIDDHA2020911}. In the COVID-19 pandemic scenario, industry and academia have begun to focus on developing remote health monitoring and disease diagnosis systems since these systems can reduce infection risk and ease healthcare burden\cite{9321459,9801732,9848442}. This paradigm can also help tackle medical deficits caused by an aging population through real-time monitoring of their physical conditions.   
	
The emergence of the Internet of Things (IoT) introduces new possibilities for remote health monitoring, early detection of chronic disease, and elderly care. IoT connects numerous devices and human beings for ubiquitous services and thus facilitates several emerging applications, such as smart factories\cite{9732420}, smart transportation\cite{7123563}, and smart home \cite{9178767}. The integration of medical devices within IoT is known as the Internet of Medical Things (IoMT) or Internet of Health Things (IoHT)\cite{9298452}. IoMT refers to the network of these medical devices connected to the internet, which can monitor vital signs, collect patient physiological data, and track medication adherence\cite{10251435}. Medical devices include wearable fitness trackers, smart watches or phones, low-power biosensors, etc.  By uninterrupted monitoring, IoMT provides doctors and caregivers with more comprehensive information about their conditions, realizing early detection of potential health issues\cite{dong2020edge}. IoMT can also integrate medical history data, such as medications, allergies, and laboratory test results, enabling health providers to make more informed treatment decisions\cite{dong2020edge}. Additionally, IoMT can help individuals make informed decisions about their health and fitness goals\cite{10251435}. Overall, IoMT is poised to revolutionize the healthcare industry, providing early prediction of potential health problems, offering patients more personalized care, and improving the quality of medical services. 
	
IoMT will facilitate several potential healthcare applications, such as real-time health monitoring, telesurgery, holographic communication, and so on\cite{8404033}. To this end, the future IoMT should have several characteristics. Firstly, the network must have stringent and real-time transmit performance since most medical applications, such as remote surgery, are time-sensitive\cite{7934392}. Secondly, the network must have the ability to support high connectivity levels to cope with the rapid increase of patients with chronic diseases. Thirdly, medical data must be delivered securely since personal health data remains private\cite{ahmed2022physical}. However, IoMT does not meet these requirements, especially low latency communication and high connectivity levels. Emerging wireless communication techniques in the fifth generation and beyond (5GB) provide a viable solution to address these challenges. For example, integrating massive multiple-input multiple-output (MIMO) and edging computing techniques reduces data transmission and processing latency\cite{9174777,8974591}. Non-orthogonal multiple access (NOMA) allows many users simultaneously to access a resource block, boosting connectivity levels\cite{8705340}. Physical-layer security (PLS) techniques can utilize signal propagation and channel characteristics to avoid privacy leakage. Also, compared to conventional cryptographic security, the PLS approach requires fewer overheads and computational complexity\cite{9127153}.

5GB-based solutions for the healthcare domain are still in the early stages, but many attempts and initiatives in this direction are underway. At this point, a comprehensive tutorial can help researchers build good foundations to guide actual design and developments. Unfortunately, the tutorial on this subject has not been investigated, which strongly motivates our work.  This paper aims to provide a thorough tutorial from the wireless communication perspective, elaborating key research progress, identifying the technique challenges and potential solutions, and highlighting future research directions.

\subsection{Prior Related Surveys}
Recently, many researchers have conducted several tutorials \cite{yaacoub2020securing,8936335,9794622,9963549,10066875,habibzadeh2019survey,9650515,8993839} or short magazine papers \cite{9490592,9099795,9083675} pertained to IoMT. These efforts report the technical developments and challenges from the clinic, edge computing, or privacy preservation perspectives, including network architecture, artificial intelligence (AI)-embedded IoMT, blockchain-enabled healthcare systems, etc. Next, the research scope and contributions of the existing surveys are presented. After that, the overall comparison of this work with the recent surveys is summarized in  Table~\ref{Table I} to highlight the distinctive contributions of our work.

The authors in \cite{yaacoub2020securing} comprehensively review security and privacy issues in IoMT and propose a promising solution to prevent privacy leakage risks. Their work further provides a comparative analysis of cryptographic and non-cryptographic-based solutions regarding computational complexity and required resources. Compared to \cite{yaacoub2020securing}, the work in  \cite{8936335} focuses on investigating different security and privacy requirements of data, sensors, person servers, and medical server levels in healthcare systems. Meanwhile, this work also points out several existing security schemes, such as cryptographic and biometric authentication approaches. The authors in \cite{9794622} conduct a survey about federated learning (FL) applications for privacy preservation in IoMT, where some advanced FL architectures incorporating reinforcement learning, digital twin, and generative adversarial networks are introduced. The medical data storage and exchange challenges are discussed and summarized in \cite{9963549}. To tackle these issues, a blockchain-based healthcare framework is presented, highlighting its benefits, future research challenges, and directions \cite{9963549}. The powerful synergy between AI and healthcare systems is explored in \cite{10066875} by elaborating systematic architecture, illustrating practical settings and many application examples, and identifying several promising future research directions. The authors in \cite{habibzadeh2019survey} survey three primary components in IoMT from the clinic perspective, including sensing and data acquisition, communication, and data analytics and inference. However, the communication-specific challenges and technologies are not mentioned in \cite{habibzadeh2019survey}, leaving this critical aspect unexplored.

Furthermore, several short briefs provide insights from the communication perspective \cite{9490592,9099795,9083675}. The NOMA-enabled healthcare system framework and several typical application scenarios are presented in \cite{9490592}, along with a discussion of potential research challenges and opportunities. The work in \cite{9099795} surveys the research efforts for implementing medical data processing and analysis, where cloud computing, edge computing, and AI technologies-aided IoMT are discussed to reduce processing delay. The edge computing-based healthcare system in IoMT is presented in \cite{9083675}, where the constructed system is separated into intra-wireless body area networks (intra-WBANs) and beyond-WBANs. The authors in \cite{9650515} divide the existing works on healthcare systems into two main taxonomies: patient-centric and process-centric techniques. Also, they provide edge computing-based solutions for the healthcare domain and discuss future challenges and trends. Several emerging technologies for IoMT are discussed in \cite{8993839}, including AI, edge computing, big data, and software-defined networks. 

The above papers or surveys collectively explore specific communication technologies, such as NOMA \cite{9490592} or edge computing \cite{9099795, 9083675,9650515}. However, to our knowledge, a critical gap remains: the absence of a systematic, comprehensive review addressing both the integration challenges and potential benefits of IoMT within diverse wireless communication paradigms. This unmet need motivates our present survey, which aims to: 1) provide a holistic communication-centric analysis of IoMT, and 2) provide an in-depth discussion on interoperability with current technologies.

	\begin{table*}[h]
	\caption{Summary of related surveys}
	\begin{center}\label{Table I}
		\begin{tabular}{|c||c|c|} 
			\hline
			\bf{References} &\bf{Focus issues} &\bf{Contributions}\\
			\hline
	        \makecell*[c]{\cite{yaacoub2020securing}} & \makecell*[c]{Security and privacy} & \makecell*[l]{
                $\bullet$ Reviewing and comparing the existing security solutions, including non-cryptographic and \\  ~~cryptographic measures;\\
                $\bullet$ Proposing a security approach by dividing IoMT into five layers to prevent attacks. }\\
	          \hline   
   	      \makecell*[c]{\cite{8936335}} & \makecell*[c]{Security and privacy} & 
                \makecell*[l]{$\bullet$ Discussing the security challenges, requirements, threats, and future research directions\\  ~~ about IoMT.}\\
			\hline 
	        \makecell*[c]{\cite{9794622}} & \makecell*[c]{Security and privacy} & 
                \makecell*[l]{$\bullet$ Conducting the FL applications for privacy preservation in IoMT;\\
                $\bullet$ Providing several advanced FL architectures incorporating other technologies, such as \\  ~~  reinforcement learning, digital twin, and so on. }\\
	          \hline 
           	\makecell*[c]{\cite{9963549}} & \makecell*[c]{Data storage and \\exchange} & \makecell*[l]{
                $\bullet$ Reviewing blockchain-based solutions in the healthcare domain;\\
                $\bullet$ Providing a systematic framework and analyzing such systems.  }\\
	          \hline 
                \makecell*[c]{\cite{10066875}} & \makecell*[c]{Data processing} & \makecell*[l]{
                $\bullet$ Exploring the synergy between AI and healthcare and providing a unified architecture;\\
                $\bullet$ Elaborating several application cases.  }\\
	          \hline 
                \makecell*[c]{\cite{habibzadeh2019survey}} & \makecell*[c]{Clinical practice} & \makecell*[l]{
                $\bullet$ Reviewing sensing and data acquisition, communication, and data analytics and inference.  }\\
	          \hline
                \makecell*[c]{\cite{9490592}} & \makecell*[c]{NOMA-aided IoMT} & \makecell*[l]{
                $\bullet$ Providing a unified NOMA framework for healthcare and several use cases; \\ 
                $\bullet$ Discussing potential research challenges and opportunities.}\\
	          \hline
                \makecell*[c]{\cite{9099795}} & \makecell*[c]{Data processing} & \makecell*[l]{
                $\bullet$ Reviewing the application of cloud computing, edge computing, and AI in IoMT.}\\
	          \hline
                \makecell*[c]{\cite{9083675}} & \makecell*[c]{Edging computing} & \makecell*[l]{
                $\bullet$ Discussing feasible solutions integrating edge computing for IoMT;\\
                $\bullet$ Dividing IoMT into two sub-networks, namely intra-WBAN and beyond-WBAN.}\\
	          \hline
                \makecell*[c]{\cite{9650515}} & \makecell*[c]{Edging computing} & \makecell*[l]{
                $\bullet$ Dividing the existing survey on healthcare systems into two categories, patient-centric and \\~~ process-centric systems;\\
                $\bullet$ Summarizing edge-computing-based solutions and highlighting future trends.}\\
                \hline
                \makecell*[c]{\cite{8993839}} & \makecell*[c]{Emerging technologies} & \makecell*[l]{
                $\bullet$ Exploring several emerging technologies for IoMT, including AI, edge computing, big data, \\~~ and software-defined networks;\\
                $\bullet$ Summarizing highlighting future trends.}\\
	          \hline
                \makecell*[c]{\bf{Our work}} & \makecell*[c]{Emerging commun-\\ication technologies} & \makecell*[l]{
                $\bullet$ Identifying a unified IoMT framework, elaborating several detailed application cases,\\~~presenting medical data transmission process;\\
                $\bullet$ Summarizing challenges of implementing IoMT, such as biosensor lifetime, co-channel\\~~ interference management (CCI), information security, and data processing;\\
                $\bullet$ Providing 5GB-based solutions, such as rate splitting multiple access, intelligent reflecting \\~~surface, physical-layer security, energy harvesting, and mobile edge computing;\\
                $\bullet$ Discussing further research directions.}\\
	          \hline
		\end{tabular}
	\end{center}
\end{table*}

\subsection{Scope and Contributions}The primary goal of this tutorial is to uncover the inherent connection between IoMT and emerging communication technologies. This tutorial distinguishes itself from prior surveys or publications on IoMT in several ways. Firstly, it thoroughly reviews recent research endeavors that integrate various communication technologies with IoMT. Secondly, it outlines IoMT features and common use cases, as a convenient reference for researchers and practitioners. Thirdly, it identifies several communication-related challenges and proposes solutions based on 5G technologies. Finally, it explores potential future research directions. The key contributions of this tutorial can be summarized as follows:
\begin{itemize}

\item Addressing gaps in the literature, a comprehensive review is conducted on the integration of wireless communication technologies with IoMT, diverging from previous works that focused solely on NOMA or edge computing in isolation from IoMT.

\item 
A three-tier IoMT framework is elaborated upon, presenting various application cases, including remote health monitoring systems, remote treatment and surgery, and enhanced clinic care and smart intensive care units (ICUs).
The medical data transmission process is elucidated, encompassing three communication stages: intra-WBAN, inter-WBAN, and beyond-WBAN communications, corresponding to data acquisition, upload, and delivery.

\item Several challenges in implementing IoMT from a communication perspective are summarized, covering biosensors' longevity, interference management, security and privacy concerns, and data processing. Solutions leveraging 5G technologies, such as novel multiple access schemes, Intelligent Reflecting Surfaces (IRS), physical-layer security, and Mobile Edge Computing (MEC), are provided to address these challenges.

\item Major research opportunities are identified, and potential research directions are discussed.
\end{itemize}

This tutorial can serve as a valuable resource for understanding the current research contributions in this area of IoMT.

\subsection{Structure and Organization}
\begin{figure}[tbp]
		\centering
		\includegraphics[scale=0.45]{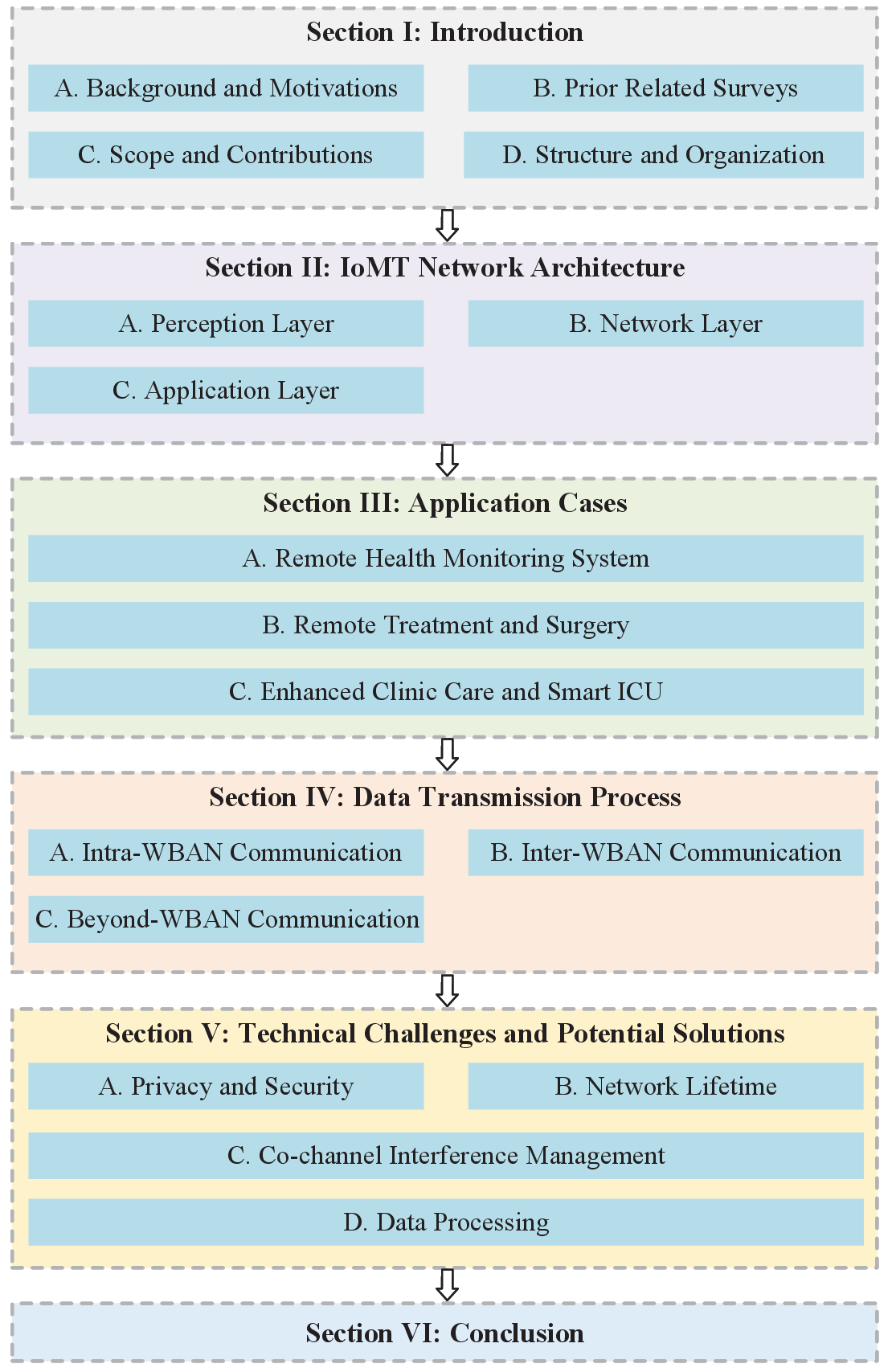}
		\caption{Organization of the present paper.}
		\label{Organization}
	\end{figure}

As illustrated in Fig.~\ref{Organization}, this paper is organized as follows. Section \ref{Section II} elaborates on three-tier network architecture and highlights their roles and functions in healthcare applications. Section \ref{Section III} and  \ref{Section IV} provide typical application cases and the medical data transmission process, respectively. Section \ref{Section V} summarizes major research challenges presenting IoMT and provides several visible solutions from the communication aspects. Also, we describe the future research directions in this section. Section \ref{Section VI} concludes this work. Some important acronyms are given in Table \ref{Table II}

	\begin{table}[]
	\caption{Summary of Important Acronyms}
	\begin{center}\label{Table II}
		\begin{tabular}{|c||c|} 
        \hline
		\bf{Acronym} &\bf{Definition}\\
        \hline 
        AI & Artificial intelligence \\
        \hline
        ECG & electrocardiogram\\
        \hline 
        EEG & electroencephalogram\\
        \hline
        FL & Federated learning\\
        \hline
        ICU & Intensive care unit\\
        \hline
	    IoMT & Internet of medical things\\
        \hline
        IRS & Intelligence reflecting surface\\
        \hline
        MIMO & Multiple-input multiple-output\\
        \hline 
        NOMA & Non-orthogonal multiple access\\
        \hline
        OMA & Orthogonal multiple access \\
        \hline 
        PLS & Physical-layer security\\
        \hline
        RSMA & Rate splitting multiple \\
        \hline
        SaO2 & Oxygen saturation\\
        \hline
        SDMA & Space division multiple access\\
        \hline
        WBAN & Wireless body area network\\
        \hline
       5GB & fifth generation and beyond\\
	    \hline
		\end{tabular}
	\end{center}
\end{table}

\begin{figure}[tbp]
		\centering
		\includegraphics[scale=0.7]{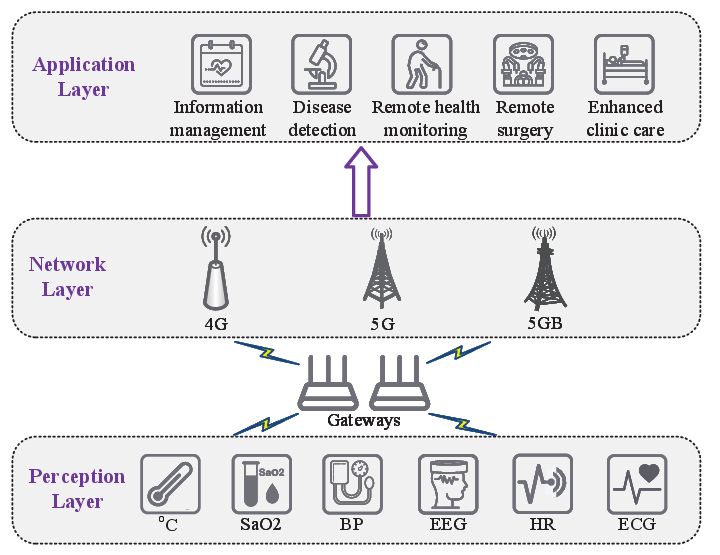}
		\caption{A three IoMT network architecture.}
		\label{IoMT architecture}
	\end{figure}

The IoMT should be competent at networking billions of diverse devices over the internet, so a flexible architectural pattern is crucial. Recently, many research scholars have proposed several IoMT architectures. These efforts mainly focus on network architecture in cloud computing or MEC\cite{9650515,10066875,9328531}. The MEC-enabled IoMT architecture comprises the edge devices layer, edge servers layer, edge clouds layer, and cloud servers layer. However, this type of architecture is specialized instead of a general framework. Driven by the traditional IoT architecture\cite{9427249,KASSAB2020102663}, this tutorial presents the most fundamental architecture, which comprises three key layers: the perception layer, the network layer, and the application layer, as shown in Fig.~\ref{IoMT architecture}. This section briefly introduces these layers, highlighting their roles and functions in the context of healthcare applications.
\section{IoMT Network Architecture} \label {Section II}
\subsection{ Perception Layer}
The perception layer, sometimes called the sensing layer, forms the foundation of the IoMT architecture\cite{9983826}. The significance of the perception layer lies in its ability to capture real-time medical data from the physical environment, transform it into meaningful information, and then transmit the information to the network layer\cite{liu2020internet}. Specifically, IoMT devices deployed in this layer can detect and measure patient vitals and other parameters, such as temperature, blood oxygen saturation (SaO2), blood pressure (BP), electroencephalogram (EEG), heart rate (HR), and electrocardiogram (ECG), and various others\cite{10066875}. IoMT can provide a holistic view of patients' conditions by combining data from multiple sensors or sources. This contextual information enhances intelligence and decision-making capabilities.  Thus, the perception layer acts as the interface between the physical and digital realms. It converts analog signals from sensors into digital data formats that can be processed and analyzed by higher layers\cite{liu2020internet}. This conversion enables interoperability, allowing different devices and systems to communicate and exchange information seamlessly.

To transfer the collected data to the network layer, the perception layer contains a short-range communication system with the gateway\cite{10251435}. This system can use a variety of access methods to connect the medical data to the network layer through short-distance data transmission technology, such as ZigBee, Wi-Fi, Bluetooth, etc\cite{8404033,8993839,dizdarevic2019survey}. With the intelligence of sensors, the perception layer enables localized decision-making and real-time responsiveness. Instead of sending all the medical data to the cloud or central server, the perception layer filters and processes part of the information locally\cite{9099866}. This reduces latency, conserves network bandwidth, and enables faster and more efficient decision-making.

Another important aspect of the perception layer is data pre-processing. Raw medical data collected by sensors usually contains noise, redundancies, or irrelevant information. The perception layer performs data cleansing, filtering, and aggregation to ensure that only relevant and high-quality data is transmitted to the upper layers\cite{8993839}. This pre-processing helps to optimize resource usage, reduce storage requirements, and improve overall system performance. 

\subsection{ Network Layer}
The network layer, e.g., the communication layer, can be divided into two sub-layers: the network transmission sub-layer and the service sub-layer\cite{9869705,9099795}. The network transmission sub-layer is a critical component of the IoMT, equivalent to the nerve center and brain of human beings. This sub-layer provides data with routing channels to transmit it in packets through the network. To this end, it uses the mobile communication network to enable medical devices to access the network and exchange data with other systems\cite{9099795}. Meanwhile, it implements various security mechanisms, such as encryption, authentication, and access control to safeguard sensitive information\cite{9011598}. As a result, the network layer ensures seamless, secure communication and connectivity between devices and systems. In addition, IoMT often involves real-time applications and critical data, such as remote patient monitoring and telesurgery. The network transmission sub-layer is crucial in managing such applications' quality of service (QoS) requirements. It prioritizes traffic based on its importance, allocating bandwidth and network resources accordingly. By ensuring adequate QoS, the network layer enables the uninterrupted delivery of healthcare services.

 The service sub-layer realizes the integration of heterogeneous networks and many data formats\cite{9099795}. Meanwhile, it is responsible for establishing a service support platform with an open interface for application services. This feature allows different manufacturers to develop various relevant applications. Consequently, the network layer has scalability and interoperability to accommodate the ongoing expansion of IoMT. It also adapts to accommodate the growing number of connected devices, ensuring the seamless integration of new technologies and healthcare solutions.

 \subsection{ Application Layer}
The application layer serves as the interface between users and healthcare services, which aims to enhance patient care, improve diagnostic accuracy, and optimize treatment strategies. Specifically, it utilizes resources in the cloud platforms to store and analyze medical data from the perception layer for specific operations and services. This layer enables various healthcare applications, including remote health monitoring, telemedicine or telesurgery, predictive analytics, disease detection, and personalized healthcare management. These applications facilitate data analysis, visualization, and decision-making processes, providing healthcare professionals and patients with actionable insights.

\section{Application Cases}\label{Section III}

  	\begin{figure}[tbp]
		\centering
		\includegraphics[scale=0.6]{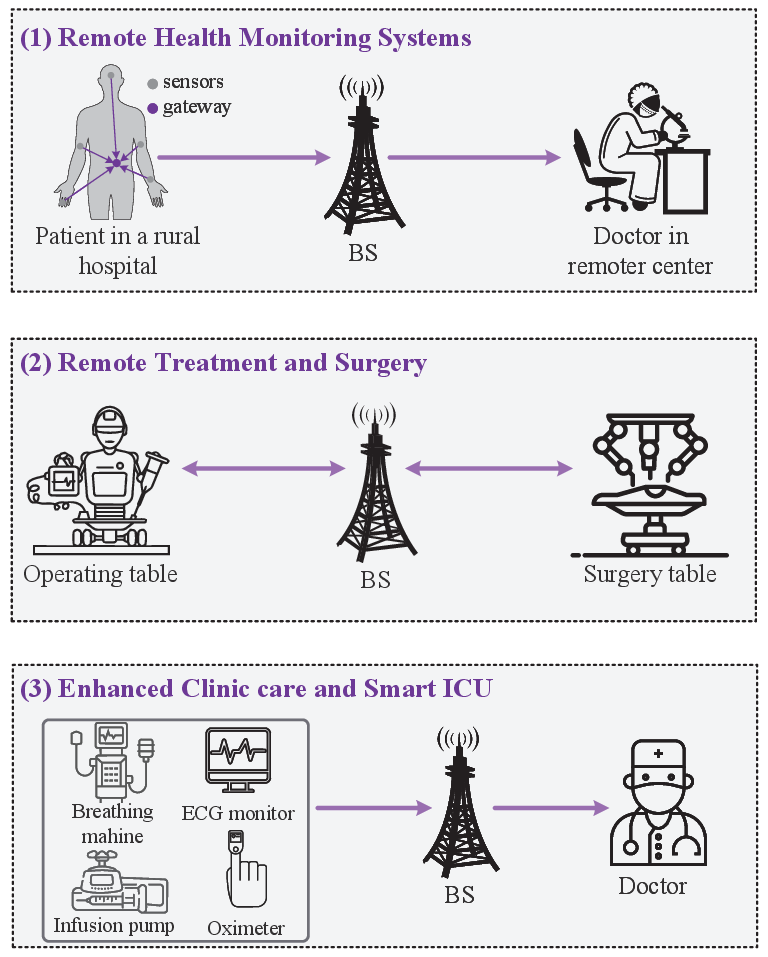}
		\caption{Several application cases in IoMT.}
		\label{fig: Application cases}
	\end{figure}
During the outbreak of the communicable disease COVID-19, heading to healthcare centers poses threats to public health.   To avoid the spread of the epidemic, patients with COVID-19 need to be isolated during this crucial timeframe\cite{KAPOOR202061}. Besides, frequent hospital visits are challenging for rural patients, and even sometimes become impossible for the elderly, who experience at least one severe chronic disease. These realities urgently call for a new paradigm to achieve remote health monitoring, disease diagnosis, and treatments. The IoMT has emerged as a game changer, expected to revolutionize the healthcare industry and ease the economic strain\cite{2020Application}.  This section provides several significant applications of IoMT in the healthcare domain and the latest research advances, including remote health monitoring systems, remote treatment and surgery, enhanced clinic care, and smart ICU. The major applications are summarized as Fig.~\ref{fig: Application cases}. Note that many applications can use remote health monitoring systems to implement their functions, such as individual healthcare, elderly care, patient healthcare, disease detection, and preventing the spread of epidemics. The nature of these applications is analogical, so they are classified into remote health monitoring systems.

\subsection{Remote Health Monitoring Systems}
The most fundamental application of IoMT is remote health monitoring systems\cite{9121312}, which utilize biosensors to measure physiological parameters.  Systems can acquire comprehensive vital sign parameters by deploying multiple biosensors on patients. Then, these data can be thoroughly analyzed to improve healthcare or predict hidden diseases. Consequently, through continuous medical data acquisition and analysis, remote health monitoring systems can provide differentiated services for different groups or medical scenarios. Three major service cases are discussed below.

\vspace{3mm}\emph{1) Individual Healthcare}

Many diseases may be completely reversed by identifying some early symptoms and then adjusting lifestyle. For example, more exercise is likely to eliminate fatty liver disease. Without intervention, long-term fatty liver may cause irreversible cirrhosis or even liver cancer\cite{6387584}. However, it is difficult for non-specialists to realize the underlying symptoms in the early stages of several chronic diseases. Fortunately, remote health monitoring systems can help individuals gain insights into their potential health issues without disrupting their lifestyles\cite{9083675}. It contributes to personal overall well-being due to three benefits. Firstly, monitoring systems allow individuals to track their health parameters, including physical activity\cite{2021Performance,8307476}, sleep patterns\cite{2017Vital}, and other wellness-related metrics. As a result, individuals can gain insights into their health status and make informed decisions regarding lifestyle choices and healthcare routines. Secondly, monitoring systems can generate personalized recommendations by collecting and analyzing physiological data. Users can receive tailored advice on nutrition and exercise modifications. Thirdly, monitoring systems encourage self-awareness and self-management. Individuals can actively participate in healthcare by setting health goals and tracking progress.

Additionally, compared with the general population, the elderly and patients with chronic diseases need more frequent medical examinations. This is challenging for such groups, especially for elderly living alone or disabled people. Remote health monitoring systems provide possibilities to track their health parameters continuously. By providing real-time data on vital signs, medication adherence, and symptom tracking, systems empower patients to manage their conditions actively \cite{habibzadeh2019survey}. They can also receive personalized reminders for medication schedules and doctor's appointments, ensuring better disease management and improved treatment outcomes\cite{9083675}. For example, remote healthcare systems effectively monitor the movement symptoms associated with Parkinson's disease and Huntington's disease\cite{9075733}. In addition, systems can monitor any fluctuations or deviations from the normal range, enabling timely intervention and preventive measures\cite{9348125,9398755}. As for the elderly living alone, systems allow caregivers or their families to remotely monitor their health conditions and receive notifications in case of emergencies\cite{2020Designing}. This helps in early identification of any health issues or emergencies.

Numerous works have been conducted to promote the application of healthcare systems in recent years. For example, the authors in \cite{9326425} develop an in-home healthcare monitoring platform that dynamically accommodates multiple sensors and thus avoids dependencies on physical ports. Monitoring systems are deployed to steer the health status of athletes during physical training and soldiers in various military operations \cite{2021Performance,8307476}, which can effectively avoid the potential damage caused by high-intensity exercise. It can also monitor vital signs during sleep\cite{2017Vital}, classify sleep stages, and detect sleep disorders, such as obstructive sleep apnea\cite{2017Validation}. In addition to sleep quality, high stress levels can cause serious physical and mental illness, even depressive disorder, in the absence of intervention in time.  Some applications attempt to monitor various types of stress and provide recommendations on a mobile device, such as post-traumatic stress disorder \cite{mcwhorter2017wearable,10027037}.

\vspace{3mm}\emph{2) Disease Detection} 

As discussed earlier, remote health monitoring systems can integrate with advanced technologies, such as AI and machine learning algorithms, to analyze the collected data \cite{8464257,8768883,8885533}. These technologies can analyze vast amounts of data and identify patterns indicating specific diseases or health issues. By leveraging these capabilities, monitoring systems can assist healthcare professionals in making accurate and timely diagnoses. For example, AI-enabled healthcare systems can use EEG data to detect and diagnose various neurological conditions\cite{10066875}. Non-invasive sensors based on sweat can be deployed to measure blood glucose levels and then predict diabetes and related conditions\cite{2021Touch}. Also, healthcare systems can utilize sensing devices to achieve early detection of dementia \cite{2016Early}. This early detection facilitates prompt medical intervention and potentially better treatment outcomes.

\vspace{3mm}\emph{3) Preventing Outbreaks of Epidemics }

When communicable diseases break out, such as COVID-19, a significant way to slow down the spread of the virus is to maintain social distancing \cite{2020The}. Thus, for patients with infectious diseases, face-to-face diagnosis and treatments are impracticable since the risk of infection will increase. Compared to the traditional medical approach, remote healthcare technologies exhibit important advantages in epidemic prevention\cite{9309371,10066875}. These technologies allow doctors and caregivers to remotely monitor their health conditions and provide remote healthcare services and telemedicine consultations. This prevents the spread of the epidemic, protects public health, and provides accurate medical services to patients. In addition, monitoring systems enable individuals to quickly identify abnormal readings that may indicate the characteristics of infectious diseases. Early detection allows for immediate isolation, thereby blocking the source of infection. This is also an effective factor to fight the outbreak of infectious diseases.

\subsection{Remote Treatment and Surgery}
In the age of 5GB communication technology, IoMT facilitates real-time communication and collaboration between patients, caregivers, and healthcare professionals via a wireless network\cite{9330766}. Healthcare providers can adjust treatment plans and make more informed treatment decisions by sharing health data and remote consultations.  This also allows patients to receive medical advice from top professions and follow-up care from their homes, reducing the need for frequent hospital visits and improving healthcare accessibility\cite{9983826}. Additionally, IoMT provides a feasible and reliable avenue for performing surgical procedures remotely by specialists through high-speed connectivity\cite{8993839}. Remote surgery enables the seamless connection between patients and surgeons by utilizing advanced wireless networks and surgical robots.  It is similar to a master-slave system, where surgeons can remotely manipulate surgical instruments with precision and accuracy at the operating site\cite{s22124577}.  The major components of the surgical robots are three-dimensional cameras with high resolution and trustworthy touch sensors\cite{9983826}. The real-time transmission of high-definition video and imaging allows for a comprehensive view of the surgical site. Based on this,  surgeons can perform complex procedures on patients in different geographic locations. In addition to expanding access to specialized surgical expertise, telesurgery reduces the risk of infections since the surgery is contactless.

\subsection{Enhanced Clinic Care and Smart ICU}
In practical clinic applications, IoMT enables the integration of electronic health records and clinical decision support systems\cite{s131217472}. This allows healthcare providers to access comprehensive patient information, including medical history, test results, and medication records. With these data, healthcare professionals can collaborate with colleagues or remote top experts to make treatment and care schemes. During the treatment, smart medication dispensers and connected pill bottles can remind patients to take their medications at the prescribed times. Systems can also alert healthcare providers or caregivers if missed doses or medication errors occur. By promoting medication adherence, IoMT improves treatment efficacy and reduces medication-related complications. The ICU requires more stringent clinical care than the general ward. To this end, vital signs monitoring and collecting equipment in the ICU can be connected to the network, such as the breathing machine, ECG monitor, infusion pump, and oximeter. These collected medical data will be converted into a standard format and sent to the data acquisition platform. Then, ICU signs monitoring,  electronic medical records, and other upper intelligent applications achieve automatic electronic medical record entry and real-time monitoring to ensure the safety of clinical care\cite{9786791,9351912}.
IoMT enhances care coordination, reduces medical errors, and enables more accurate clinical decision-making. 

\section{Data Transmission Process}\label{Section IV}
 	\begin{figure}[tbp]
		\centering
		\includegraphics[scale=1.1]{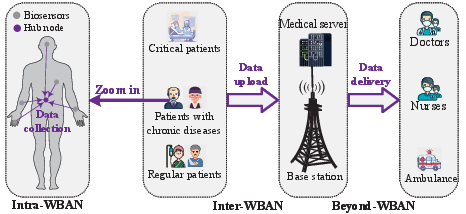}
		\caption{Data transmission process of remote health monitoring systems.}
		\label{fig: Data transmission}
	\end{figure}
Implementing the above IoMT applications is inseparable from high-speed data transmission and processing. With high-speed data transfer and processing, medical devices like wearable monitors and smart implants can seamlessly relay patient data to healthcare professionals, enabling swift diagnosis and personalized treatment plans. Moreover, remote treatments require more stringent transmission performance to smooth video conferencing and virtual consultations between physicians and patients. In this section, remote health monitoring systems are used as an example to illustrate the data transmission process. Following the approach of \cite{movassaghi2014wireless}, the monitoring systems are categorized into three sub-networks: intra-Wireless Body Area Network (intra-WBAN), inter-WBAN, and beyond-WBAN, as depicted in Fig.~\ref{fig: Data transmission}. These sub-networks respectively handle data acquisition, upload, and delivery.

\subsection{Intra-WBAN Communication}
Intra-WBAN, a physical network system with data acquisition and transmission functions, is fundamental for achieving remote health monitoring, disease detection, and accurate treatment decision-making\cite{9241882}. An intra-WBAN comprises a group of wearable or implanted biosensors and an on-body hub node (e.g., smartphone, tab, or laptop). A biosensor includes a biological component, such as an enzyme or antibody, and a transducer. The former is used for recognizing a specific biological molecule or analyte, while the latter is used to convert the binding event into an electrical signal. Biosensors can be designed to detect various signals and analytes, including ECG, EEG, SaO2, blood pressure, glucose, cholesterol, hormones, bacteria, and viruses\cite{aledhari2022biomedical}. Intra-WBAN utilizes these biosensors to detect and measure various biological or chemical reactions, collecting personal physiological parameters. These parameters are transmitted to the hub node by using short-distance communication technologies\cite{askari2021energy,ning2020mobile}. The more technologies supported by intra-WBAN, the easier for users to be integrated into applications. The communication is facilitated using IEEE 802.15.6\cite{astrin2012ieee}, which leverages existing frequency bands authorized by national medical authorities. Note that the coverage distance of the WBAN devices in transmitting and detecting signals is limited to within or around the human body. 

This layer is attracting the attention of researchers since it is the basis of developing IoMT applications. Diverse works have investigated intra-WBAN communications to enhance data transmission performance \cite{9241882,ashraf2022energy,7934392,9292984,9115049,9434926} or ensure privacy security\cite{8756092,jabeen2021survey}. For example, the authors in \cite{9241882} consider the co-frequency interference management between coexisting WBANs in dense environments. A spectrum allocation scheme based on machine learning is proposed to suppress interference levels to achieve intelligent partition. A dynamic channel allocation algorithm under IEEE 802.15.6 standard is designed to exploit the benefit of the polling access mechanism \cite{ashraf2022energy}. A price-based opportunistic data dissemination method is studied to maximize energy efficiency, considering user mobility \cite{7934392}. To enhance the reliability of emergency-critical data transmission, the authors in \cite{9292984} propose a resource scheduling strategy and a deep reinforcement algorithm to solve the formulated non-convex problem. In \cite{9115049}, the device-to-device (D2D) communication mode connects biosensors to the hub node to avoid collisions among WBAN users. To improve system reliability, WBAN can integrate buffer-aided relaying to reduce the power consumption of the implant biosensors \cite{9434926}. An efficient pair-free aggregate signature scheme to proposed to ensure the privacy and security of medical data\cite{8756092}. However, user mobility in intra-WBAN remains largely unexplored, with the exception of\cite{7934392}.
 
\subsection{Inter-WBAN Communication}
This communication layer aims to connect the hub nodes and the base station (BS) or access point (AP) via various networks. As a result, hubs forward medical data to the BS with an edge server for further data processing. When detecting unusual signals, the user can be rapidly located, tracked, and provided with the most suitable medical services\cite{qiu2021computation}. Currently, the paradigms of inter-WBAN communication can be divided into two categories, namely infrastructure-based architecture and Ad-hoc-based architecture \cite{movassaghi2014wireless}. The former can be used for most IoMT applications since it facilitates dynamic deployment in a limited
space. The latter can establish a mesh construction, providing larger radio coverage since it allows multi-hop dissemination.

The research on inter-WBAN communication focuses on two aspects. First, some works aim to boost the system performance of inter-WBAN \cite{6193096,8501564}. To reduce the interference among WBAN users, the authors in \cite{6193096} propose a random incomplete coloring algorithm with low time complexity and high spatial reuse for mobile inter-WBAN. Inter-WBAN can integrate with other technologies. For example, inter-WBAN employs simultaneous wireless information and power transfer (SWIPT) and full-duplex relay to boost system performance \cite{8501564}. Second, some efforts focus on the joint design of intra-WBAN and inter-WBAN\cite{askari2021energy,ahmed2022physical,dong2020edge,ning2020mobile}. The authors in \cite{askari2021energy} propose a Walsh-Hadamard code scheme for intra-WBAN and a resource allocation algorithm for NOMA-assisted inter-WBAN. A federated learning-based PLS approach for intra-WBAN and inter-WBAN communication can protect medical data privacy \cite{ahmed2022physical}. Game theoretic solutions can minimize system-wide costs in health monitoring systems \cite{dong2020edge,ning2020mobile}, where intra-WBAN and beyond-WBAN communications are modeled as cooperative and non-cooperative games, respectively. 

\subsection{Beyond-WBAN Communication}
The design of beyond-WBAN is for use in remote areas, from the Internet to the medical server in a specific application \cite{latre2011survey}. A gateway can be used to bridge the connection between inter-WBAN and beyond-WBAN so healthcare providers can access comprehensive patient information. With these data, doctors can make more informed treatment and care decisions. However, the processed data should be transmitted to medical personnel delay-sensitively to provide timely medical service. To this end, remote health monitoring systems can integrate with advanced communication technologies\cite{tang2020uav,9181534,9814697}.  Unmanned aerial vehicles (UAV) placement optimization is studied to minimize power cost \cite{tang2020uav}. Multi-carrier NOMA is used to suppress the inter-carrier interference while the optimized system capacity and symbol error with closed-form expression are derived in \cite{9181534}. Intelligent reflecting surface (IRS) technology can be integrated to ensure privacy and security during the data transmission process\cite{9814697}.

\section{Technical Challenges and Potential Solutions}\label {Section V}
The increasing prevalence of chronic diseases has led to a rise in patients, resulting in heightened wireless data transfers within the IoMT. This surge has brought challenges primarily focused on privacy issues, network longevity, co-channel interference, and data processing. This section delves into the technical hurdles faced by IoMT applications. Moreover, potential solutions are proposed to address these challenges, leveraging advanced wireless communication technologies such as energy harvesting, rate splitting multiple access (RSMA), MEC, and IRS. Despite their promise, these advanced wireless technologies have been relatively underexplored in  IoMT. To propel IoMT forward across various applications,  futuristic visions and innovative research directions are suggested.

\subsection{Privacy and Security}
\begin{figure}[tbp]
		\centering
		\includegraphics[scale=0.9]{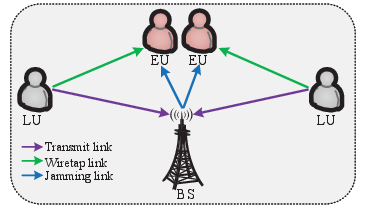}
		\caption{An artificial noise-aided uplink secure network, where EUs and LUs denote eavesdropping users and legitimate users, respectively.}
		\label{fig: AN-aided PLS}
	\end{figure}

\emph{Technical challenge:} Secure wireless networking, which ensures that only legitimate users can access the information, is of utmost importance in IoMT. This is because sensitive medical data must be personally concealed. Also, medical data leakage may cause several drastic consequences. For example, malicious parties may take control of healthcare systems and then tamper with medical data or treatment protocols, threatening patients' health and even lives\cite{9801732}. With data privacy and security accentuated in biomedical society, the assurance of security in such systems is imperative for its wide implementation and further development. Nevertheless, the inherent broadcast nature of the wireless medium incurs a myriad of potential security attacks. Traditional wireless networks utilize secret-key encryption methods to combat adversarial users, requiring higher overheads and computational complexity\cite{9127153}. The PLS approach is proposed in this context, which leverages signal propagation and channel characteristics to counter malicious attacks\cite{8543573}. Compared to conventional cryptographic security, its distinctive advantage lies in the boosted security without calling for additional overheads. Various PLS techniques have been proposed for downlink wireless networks\cite{9504435,8533374,8786136,9844767}. However, when eavesdroppers and legitimate receivers closely resemble channel characteristics, designing transmit beamforming alone may be insufficient to safeguard against privacy breaches\cite{9127153}. In such scenarios, several additional helpers, such as cooperative jammer or artificial noise (e.g.,\ref{fig: AN-aided PLS}), become extremely necessary to fortify security measures. These measures help ensure security and privacy, but they have two disadvantages. First, it weakens the information reception capacity of legitimate receivers, reducing the spectral efficiency. Second, it requires extra transmit power to design artificial noise or cooperative jammer, impairing energy efficiency. These two adverse effects urgently require novel PLS techniques to avoid information leakage.

  	\begin{figure}[tbp]
		\centering
		\includegraphics[scale=0.9]{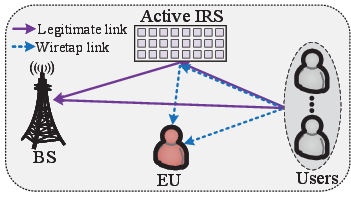}
		\caption{An IRS-aided uplink secure network.}
		\label{fig: IRS-aided PLS}
	\end{figure}
\emph{Potentail solutions:} From the information theory perspective, PLS techniques ensure that the decoding ability of legitimate users is greater than that of eavesdroppers. Therefore, there are three ways to facilitate PLS: 1) enhancing the decoding capacity of legitimate users, 2) weakening the decoding capacity of malicious users, or 3) simultaneous occurrence of 1) and 2)\footnote{In fact, using artificial noise and cooperative jamming are the second solution.}. The third approach is a more attractive solution since it is more prone to ensure communication security. IRS is a revolutionary game changer for achieving PLS\cite{10217345,9807309,9520776}. Specifically, the IRS can inherently manipulate wireless channels/radio propagation environment by adjusting its reflecting parameters. This programmability allows us to deliberately strengthen or weaken the reflected signal toward specific users in the wiretap channels\cite{9279316}. This bestows the IRS technique great potential to enhance secure communication. Fig.~\ref{fig: IRS-aided PLS} elaborates an IRS-aided PLS example. At present, two types of IRS exist: passive and active. The former lacks signal amplification and only imparts a phase change, whereas the latter, with an additional power source, can change the phase and amplify incident radio frequency signals. Passive IRS-enabled PLS approaches have recently been investigated for downlink and uplink systems\cite{10217345,9807309,9520295,9941150,10225700}. However, the double path-loss from the transmitter-IRS channel and IRS-receiver channel, also known as multiplicative fading, limits the performance of the passive IRS networks. When the direct channel link between the transmitter and receiver is unobstructed, the existing passive IRS only yields a negligible capacity gain\cite{9998527}. In response, the active IRS has emerged. The benefit of active IRS has been investigated in\cite{9998527, 9896755}, but active IRS-aided secure communication has not been explored.

Another potential candidate for securing IoMT networks is RSMA. As discussed above, RSMA splits and encodes users' messages into a common stream and multiple private streams. When the common stream cannot be decoded by eavesdroppers, it can be used for dual purposes, serving as the desired stream for legitimate users and artificial noise for eavesdroppers\cite{9721222}. Therefore, this common stream enhances legitimate users' decoding capacity while impeding eavesdroppers' information reception. RSMA-enabled PLS has been explored in \cite{9721222,10415047,li2020cooperative,10285055}, which validates that this dual-purpose property helps enhance secure communication.

 \emph{Future directions:} The IRS technique has exploited great advantages in avoiding information leakage, but several practical issues still need to be addressed. Firstly, many existing works require perfect CSI between the IRS and eavesdroppers to design IRS beamforming\cite{10217345,9807309,9520295,9941150,9721222,li2020cooperative}. If perfect CSI is unavailable, the performance gains degrade significantly. CSI is generally estimated by feedback link, but finite-rate uplink may not accurately describe the perfect CSI, especially for multi-antenna or multi-user systems\cite{8278213}. Also, the eavesdroppers may intentionally remain covert, aggravating the difficulty of estimating CSI. This calls for a robust IRS beamforming design that considers the imperfect or statistics CSI. Secondly, as the elderly population increases, IoMT inevitably presents many legitimate users and eavesdroppers. In such networks, multi-IRS deployment deserves further study since it significantly impacts system performance.

\subsection{Network Lifetime}
\emph{Technical challenge:} In the context of IoMT, intra-WBANs use lightweight biosensors to gather and transmit vital signs to an accessed hub node. Each patient must be deployed with multiple biosensors to collect more comprehensive health information. Biosensors serving the same patient are generally in mutual proximity since they are embedded in or implanted in the patient body. Meanwhile, as patients with chronic diseases and the elderly population increase, multiple users should be allowed to access the intra-WBAN simultaneously. According to the latest WBAN protocol IEEE 802.15.6, intra-WBAN needs to reach a 3-meter transmission range, 10 users in a range of 63 cubic meters, and a 10 Mbps (bit/s) transmission rate\cite{astrin2012ieee}. In a high-density deployment scenario, multiple biosensors must share the same resource block as the spectrum resource exhausts\cite{9241882}. When too many biosensors, especially for biosensors in mutual proximity,  are scheduled per time slot, severe CCI may cause the collected medical data to be outdated when it reaches the hub node. Even if it works, each biosensor will cost more transmit power to ensure the age of information (AoI). However, biosensors are mostly low battery-powered since they have a limited physical size. Also, changing batteries is time-consuming or even impossible for in-body biosensors. These practical limitations will seriously shorten the lifetime of biosensors. Intra-WBAN communication, as the first stage of IoMT applications, directly affects the performance of the entire IoMT network. If data acquisition and transmission of biosensors outage, inter-WBAN communication, beyond-WBAN communication, and data processing will be terminated. This may adversely impact timely data acquisition, transmission, and even accurate disease diagnosis\cite{8647119}. Therefore, it is imperative to prolong the lifetime of biosensors.

\emph{Potential solution:} Two intuitive ideas for extending biosensors' longevity is energy-saving and energy supply in a self-sufficient manner. Several existing approaches can reduce energy consumption and thus prolong biosensors' lifetime to a certain extent, such as advanced routing/communication protocols\cite{2022Reliable},  low-power radio transceivers, or efficient resource management algorithms\cite{7934392}.  The authors in \cite{2022Reliable} propose an efficient routing algorithm to minimize energy costs and CCI for intra-WBAN communication cooperatively.  A price-based
opportunistic data dissemination method is studied to maximize energy efficiency, considering user mobility \cite{7934392}. However, these approaches cannot fully avoid biosensor outages since they fail to provide a sustainable power supply. 

Unlike the energy-saving strategy, the other direction is seeking extra supply. The energy harvesting technique, a potentially self-powered solution, has attracted considerable attention\cite{9170604,8944276}. Specifically, it can capture the ambient radio frequency radiation and convert it into a direct current energy source through appropriate rectennas circuits. The harvested electricity is directly fed to biosensors for gathering and transmitting patient health data. As a result, energy harvesting-aided intra-WBAN can ensure sustainable network operation. The research on energy harvesting mainly focuses on two aspects. Firstly, many works develop protocols or resource allocation strategies for conventional wireless communication systems\cite{9333647,7797187,9351543}. These efforts demonstrate that the energy harvesting technique can boost performance or avoid network outages\cite{8705340}. Secondly, several works consider the circuit and rectenna design, which aim to promote the practical applications of energy harvesting\cite{9427092,10113814}. Driven by this benefit, the energy harvesting technique has recently been extended for intra-WBAN communication\cite{7425153}. For example, it can be used to minimize relay nodes' outage probability for relay-aided intra-WBAN\cite{8241755}. The authors in \cite{9954057} propose a switching strategy-based energy harvesting protocol to enhance energy efficiency. In addition, energy harvesting can be integrated with other technologies to yield greater gains in intra-WBAN further. The authors in \cite{8647119} leverage the synergy between sleep scheduling and energy harvesting for intra-WBAN, where these two approaches can save transmit power and provide sustainable power, respectively. Cognitive radio-based intra-WBAN can apply energy harvesting to enhance spectral efficiency\cite{9363909}. 

\emph{Future directions:} Several attempts have extended the energy harvesting technique to healthcare systems\cite{7425153,8241755,9954057,8647119,9363909}, but two research topics still deserve to be explored. Firstly, existing works consider energy harvesting-aided healthcare systems and design resource allocation algorithms from a theoretical perspective. However, biosensors and hubs have limited physical space in practical healthcare applications. This limits them from being equipped with large energy harvesters. Thus, one can design more advanced and refined energy harvesters from the hardware perspective. Such developments help to promote the application of the energy harvesting technique in IoMT. Secondly, compared to battery-powered IoMT, energy harvesting-enabled IoMT is more vulnerable to security threats\cite{8944276}. Conventional securing communication methods use secret-key encryption, but the overhead of sharing secret keys makes this measure challenging for IoMT applications with minimal energy supply. As such, an attractive research direction could be to exploit the PLS technique to avoid information leakage.

\subsection{Co-channel Interference Management}
\emph{Technical challenge:} The surge in patients with chronic ailments has increased wireless data exchange, complicating accurate interference management. Currently, CCI remains a major obstacle to implementing real-time  IoMT communications. To control interference levels, medical devices (biosensors, hubs, and BS) must adhere to specific protocols to govern data transmission times and methods, known as multiple access techniques. The prominent access schemes can be categorized into three types: orthogonal multiple access (fully avoiding CCI), space division multiple access (fully treating CCI as noise), and non-orthogonal multiple access (fully decoding CCI). These three multiple access schemes are described next.

The primary OMA technique comprises frequency division multiple access (FDMA), time division multiple access (TDMA), code division multiple access (CDMA), and orthogonal frequency division multiple access (OFDMA). These schemes divide total resources into multiple orthogonal resource blocks and then schedule users in orthogonal dimensions to completely avoid CCI. Their benefit is that low-complexity receivers can entirely separate the desired signal from the received message using different basis functions\cite{7676258}. Since OMA only serves one user over each resource block, its scalability is hindered by limited spectrum resources. Thus, it is only suitable for underloaded networks. However, the rapid growth in the elderly population has prominently propelled data traffic. Therefore, OMA cannot be used for IoMT applications.

SDMA scheme allows multiple users to share the same radio resources, e.g., time/frequency/code. Each receiver fully treats interference as noise and directly decodes the desired signal, so its precoder and receiver complexity are low\cite{mao2018rate}. SDMA can counter CCI by exploiting spatial multiplexing gains. For example, the multiple antenna technique creates many degrees of freedom (DoF) through transmit beamforming to neutralize CCI\cite{9835151,mao2018rate}. However, its performance is sensitive to the user channel orthogonality and strengths. Also, when too many users are scheduled per time slot, the number of data streams exceeds spatial DoF, leading to excessive interference\cite{mao2018rate}. Consequently, this will result in a plateaued achievable rate even if the transmit power can be increased infinitely. In other words, SDMA is suited to the underloaded regime, not an overloaded regime, since it requires more transmit antennas than users to manage CCI. Therefore, SDMA cannot concurrently provide unsaturated performance and high connectivity levels.

NOMA superposes users' signals in the same time-frequency resource via the power domain or the code domain, yielding respectively power-domain NOMA and code-domain NOMA\cite{7676258,8694790}. Power-domain NOMA\footnote{This survey mainly focuses on power-domain NOMA, which will be referred to simply by NOMA.} uses superposition coding at the transmitters and successive interference cancellation (SIC) at the receivers. The SIC technique enables each receiver to fully decode and remove strong interference signals before detecting its desired signal\cite{9502643}. This helps boost the system throughout and accommodate massive connectivity to some extent. Despite this benefit, NOMA presents several technical challenges. First, stronger signals must be decoded successively, so NOMA requires multi-tier SIC and stringent decoding orders. This limitation complicates the transmitter and the receiver design\cite{8294044,7676258}. Second, multi-tier imperfect SIC causes error propagation in subsequent decoding, resulting in decoding errors\cite{mao2018rate}. Third, NOMA deployment is only suitable for some special scenarios where the users' channels are sufficiently aligned and exhibit distinct disparity of user channel gains\cite{mao2018rate}, not in general scenarios. Fourth, its performance gain degrades gravely due to imperfect channel state information (CSI)\cite{9451194}. These requirements and prerequisites curtail NOMA's applicability in IoMT. 

	\begin{table*}[h]
	\caption{Comparison of different multiple access schemes}
	\begin{center}\label{Table III}
		\begin{tabular}{|c||c|c|c|c|} 
			\hline
			\bf{Multiple Access} &\bf{OMA} &\bf{SDMA}&\bf{NOMA}&\bf{RSMA}\\
			\hline
	        \makecell*[c]{\bf{CCI management strategy}} & \makecell*[c]{Fully avoid CCI} & \makecell*[c]{
                Fully treat CCI as noise }& \makecell*[c]{
                Fully decode CCI }& \makecell*[c]{
                Partially decode CCI and\\ partially treat CCI as noise }\\
                \hline 
           	\makecell*[c]{\bf{Receiver complexity}} & \makecell*[c]{Low} & \makecell*[c]{
                Low}& \makecell*[c]{
                High}& \makecell*[c]{
                High}\\
                \hline 
           	\makecell*[c]{\bf{User deployment scenario}} & \makecell*[c]{Any channel direction \\and any difference in \\channel gain} & \makecell*[c]{
                (semi-)orthogonal channels\\ with similar channel strengths }& \makecell*[c]{
                 Good alignment of user \\channels and distinct disparity\\ in user channel gains}& \makecell*[c]{
                Any channel direction \\and any difference in \\channel gain}\\
	          \hline 
           	\makecell*[c]{\bf{Suitable network load}} & \makecell*[c]{Underloaded network} & \makecell*[c]{
                Underloaded network }& \makecell*[c]{
                Overloaded network}& \makecell*[c]{
                Suited to any network load}\\
	          \hline
                \makecell*[c]{\bf{Achievable rate regime}} & \makecell*[c]{Small} & \makecell*[c]{
                Small}& \makecell*[c]{
                Medium}& \makecell*[c]{
                High}\\
	          \hline
		\end{tabular}
	\end{center}
\end{table*}

\emph{Potential solution:} As discussed above, the hitherto multiple access schemes lack flexibility in managing CCI, resulting in subpar performance. As a result, neither of them can support massive connectivity in time-sensitive IoMT applications. To provide medical service in the shortest time, IoMT requires a flexible CCI management approach. RSMA, a more general and powerful multiple access framework, provides greater flexibility to attack CCI, allowing for improved capacity regions\cite{10038476,9942944,9573421,9665262}. Its main idea is to split each message into multiple sub-messages, which are transmitted by using superposition coding at the transmitters and decoded using SIC at the receivers. It is worth noting that downlink and uplink RSMA have different encoding and decoding principles. Specifically, for downlink RSMA communications, the transmitter splits each message into a common part and a private part, encodes the common parts into one common stream using a common codebook, and each private part into a separate private stream. Then, the common stream is superimposed on the top of private streams. On the receiving side, each user first decodes the common stream, removes it via SIC, and then detects its desired private stream. For uplink RSMA communications, each transmitter splits its message into multiple parts and then encodes them into multiple independent streams with separate transmit power. The receiver performs SIC decoding to sequentially retrieve each stream and reconstruct the original message\cite{9257190}. However, compared to NOMA, uplink RSMA allows flexible decoding order, contributing to efficient CCI management. This new transmit scheme allows RSMA to fully exploit the available time, frequency, power, and spatial domains, yielding the optimal DoF in the downlink and capacity region in the uplink\cite{10038476}. By optimizing power allocation parameters and decoding order, RSMA decodes partial CCI and treats the remaining portion as noise, encompassing OMA, SDMA, and NOMA as special cases. Compared to the existing multiple access schemes, RSMA brings several additional benefits. First, it can be operated in general scenarios where the user channels are neither orthogonal nor aligned and exhibit similar or diverse strengths. Second, it can be used for underloaded networks and overloaded networks. Third, it presents better robustness to imperfect CSI conditions and high mobility scenarios\cite{9932426}. Therefore, RSMA is particularly suitable for time-sensitive IoMT applications without interrupting patients' daily lives. The comparison between RSMA and existing multiple access schemes is summarized in \ref{Table III}.

The idea of rate splitting dates back to the early work on the two-user interference channel by Han and Kobayashi\cite{1056307}. The authors in \cite{1056307} develop the rate splitting strategy to establish a new achievable rate region. However, as the building block of RSMA, RS is motivated by recent works, which proved its benefits in multi-antenna systems and overloaded regimes. The benefits include higher energy efficiency\cite{10032267,9693949,10269133,10208133}, higher fairness rate\cite{9991090,10208133}, better robustness\cite{9445019,9790069}, and higher DoF\cite{7434643} over SDMA and NOMA. These excellent works focus on downlink RSMA systems. Fortunately,  several works show that uplink RSMA with suitable decoding order can yield the optimal rate region without using time-sharing\cite{rimoldi1996rate, mao2022rate}. Nevertheless, existing uplink RSMA systems only consider simple user deployment scenarios. For example, the authors in \cite{9755045,9676684,xu2022rate,yue2022ergodic} investigate the outage probability for two-user uplink systems. Reference \cite{yang2020sum} considers the sum rate maximization problem for simple multi-user systems. Currently, uplink RSMA has not been considered for IoMT, except \cite{10251435}. The authors in \cite{10251435} jointly design decoding order, power allocation, and receiver beamforming to minimize the time cost.

\emph{Future directions:} Currently, the research outcomes of RSMA mainly lie in the downlink systems\cite{10032267,9693949,10269133,9991090,9445019,9790069,7434643,10312769}. However, in IoMT applications, the collected medical data must be transmitted to a central node, predominantly through uplink communications. Therefore, exploring uplink RSMA for general IoMT network deployment is imperative. Three potential avenues are highlighted below. According to uplink RSMA schemes, the receiver must be equipped with multi-tier SIC. Imperfect SIC may cause decoding errors, yielding error propagation in the subsequent decoding process. This means that if one signal is decoded incorrectly, so does the remaining part. Also, hub nodes are difficult to equip multi-tier SIC due to the limited physical size. Thus, it is necessary to develop novel uplink RSMA schemes to reduce the required SIC layers. Secondly, different from downlink RSMA, uplink RSMA has a flexible decoding order, so designing the corresponding resource allocation algorithms is required. Thirdly, the existing RSMA works mainly focus on static systems. However, remote health monitoring or diagnosis systems aim to optimize patient outcomes without interrupting their daily life. As such, an intriguing exploration direction is to develop resource allocation algorithms tailored to the dynamics of user mobility scenarios. This area of inquiry has immense potential to improve patients' happiness considerably.
 
\subsection{Data Processing}
\emph{Technical challenge:} Traditional IoMT uses cloud computing to address computing, storage, and network management problems centrally, transferring the collected medical data to the cloud computing center\cite{8404033}. The multi-cloud framework can be designed and utilized to reduce information processing and storage costs, which has broad applicability and scalability\cite{8862848}. Cloud computing has gradually evolved into a powerful and mature network service platform\cite{7912249}. However, its inherent limitation is the long propagation distance from the mobile user to the remote cloud center, which causes a high transmit delay for mobile applications. Meanwhile, with the surge in patients and elderly populations, the number of medical devices connected to IoMT is rapidly increasing. A mountain of medical data is generated every second and needs to be offloaded to the remote cloud center. Also, the processed data should be fed back to patients or caregivers in time. Both two data transmission phases put a serious burden on access links. Even dark fiber with tens of Gbps capacity could be prone to congestion and thus deteriorate transmission delay\cite{peng2016recent}. As a result, cloud computing-enabled IoMT is no longer sufficient to complete some delay-sensitive and computation-intensive computing tasks.

\emph{Potential solution:} One of the promising solutions is the MEC technique, which offloads the raw physiological data and computation tasks to edge servers for preliminary processing\cite{8016573}. The disease diagnosis and in-depth data processing or sharing can be carried out on a cloud server. In other words, MEC migrates computing, storage capacities, and resources to the network's edge, bringing these task executions closer to the source of the data\cite{9083958}. Therefore, MEC-enabled IoMT reduces the network burden greatly and thus provides numerous benefits, such as shortened transmission distance and delay, improved quality of experience, and energy efficiency\cite{9083958}. 

Recently, the application of MEC-enabled IoMT has been gradually explored in\cite{9083675,qiu2021computation,bishoyi2021enabling,ning2020mobile,9186655,9480807,9184073,10634173}. For example, the authors in \cite{qiu2021computation} utilize MEC to alleviate the burden of the cloud center, where the resultant non-linear and non-convex problem is decoupled into three independent sub-problems. Game theoretic solutions can be used to make suitable offloading decisions in MEC-enabled IoMT, which models this problem as a Stackelberg game\cite{bishoyi2021enabling}. Then, an alternating direction method of multipliers (ADMM)-based distributed algorithm is proposed to solve the Stackelberg game and reach the Stackelberg equilibrium\cite{bishoyi2021enabling}.  References \cite{9083675} and \cite{ning2020mobile} consider intra-WBAN and inter-WBAN, where these two communication phases are modeled as cooperative and non-cooperative games, respectively. Besides, task offloading can be modeled as an adversarial multi-armed bandit problem\cite{9186655}. Reinforcement learning can be used to make offloading decisions to minimize latency and energy consumption\cite{9480807}. In MEC-enabled IoMT, user mobility may cause uncertain connection states between them to edge servers. The authors in \cite{9184073} investigate the impact of this uncertainty on offloading decision-making and resource allocation. Reference \cite{10634173} proposes the uplink and downlink RSMA transmit scheme for fog computing-enabled IoMT. The results show that the proposed transmit scheme can significantly reduce overall
time cost encompassing task offloading, data processing, and result feedback\cite{10634173}.

\emph{Future directions:} MEC helps ease backhaul burden and reduce transmit delay by processing data in edge nodes. In the context of IoMT applications, the computing power of hub nodes is limited due to the limited physical size. As a result, part of the collected medical data still needs to be transferred to the cloud computing center. However, the existing MEC-enabled IoMT mainly adopts SDMA or NOMA. As discussed earlier, these two multiple-access technologies cannot efficiently manage CCI. This compromises transmission delay and is incompatible with providing medical services in the shortest time. Therefore, the MEC can be further explored in the IoMT enabled by RSMA. Such studies would be of great benefit to both patients and medical practitioners.

\section{Conclusion}\label{Section VI}
This survey offers a comprehensive overview and research outlook on the IoMT from a communication perspective. It encompasses a thorough literature review of recent research endeavors in IoMT, highlighting this survey's unique contributions. Beginning with an introduction to the three-tier architecture of IoMT and each tier's primary functions, the survey examines four application cases: remote health monitoring systems, remote treatment and surgery, enhanced clinic care, and smart ICU. Using the remote health monitoring system as a focal point, the survey elaborates on the medical data transmission process, encompassing intra-WBAN, inter-WBAN, and beyond-WBAN communications. Building upon this foundation, the survey outlines four challenges in implementing IoMT and proposes corresponding solutions from a wireless communication perspective. Additionally, it identifies several research directions aimed at further facilitating IoMT development. This comprehensive tutorial on IoMT aims to serve as a valuable reference and guide for deeper investigations into IoMT.

	\ifCLASSOPTIONcaptionsoff
	\newpage
	\fi
	
	\bibliographystyle{IEEEtran}
	\bibliography{references}
	
\end{document}